\documentclass[oldversion]{aa-v6.1}

\newcommand{\beal}{\begin{align}}
\newcommand{\id}{{\,\rm d}}
\newcommand{\Abst}[1]{\,#1}
\newcommand{\pot}[2]{#1 \times 10^{#2}}
\newcommand{\lesssim}{\mathrel{\hbox{\rlap{\hbox{\lower4pt\hbox{$\sim$}}}\hbox{$<$}}}}
\newcommand{\gtrsim}{\mathrel{\hbox{\rlap{\hbox{\lower4pt\hbox{$\sim$}}}\hbox{$>$}}}} 
\newcommand{\nmax}{n_{\rm max}} 
\newcommand{\nsplit}{n_{\rm split}}

\newcommand{\kB}{k}
\newcommand{\me}{m_{\rm e}}
\newcommand{\Ne}{N_{\rm e}}
\newcommand{\Tg}{T_{\gamma}}
\newcommand{\Te}{T_{\rm e}}

\newcommand{\pd}{\partial}
\newcommand{\pAb}[2]{\frac{\displaystyle\pd #1}{\displaystyle\pd #2}}

\usepackage{amsmath}
\usepackage{natbib}
\usepackage{epsfig}
\usepackage{txfonts}

\begin{document}
\titlerunning{Free-bound emission from cosmological hydrogen recombination} 
\title{Free-bound emission from cosmological hydrogen recombination}

\author{J. Chluba\inst{1} \and R.A. Sunyaev\inst{1,2}}
\authorrunning{Chluba \and Sunyaev}

\institute{Max-Planck-Institut f\"ur Astrophysik, Karl-Schwarzschild-Str. 1,
85741 Garching bei M\"unchen, Germany 
\and 
Space Research Institute, Russian Academy of Sciences, Profsoyuznaya 84/32,
117997 Moscow, Russia
}

\offprints{J. Chluba, 
\\ \email{jchluba@mpa-garching.mpg.de}
}

\date{Received / Accepted}

\abstract{
In this paper we compute the emission coming from the direct recombination of
free electrons to a given shell ($n\geq 2$) during the epoch of cosmological
hydrogen recombination. This contribution leads to a total of {\it one} photon
per recombined hydrogen atom and therefore a $\sim 30-88\,\%$ increase of the
recombination spectrum within the frequency range $1\,\text{GHz}\leq
\nu \leq 100 \,\text{GHz}$.
In particular the Balmer-continuum emission increases the distortion at
$\nu\sim 690\,$GHz by $\sim 92\%$.
With our 100 shell calculations for the hydrogen atom we find that a total of
$\sim 5$ photons per hydrogen atom are emitted when including all the
bound-bound transitions, the 2s two-photon decay channel and the optically
thin free-bound transitions.
Since the direct recombination continuum at high $n$ is very broad only a few
$n$-series continuua are distinguishable and most of this additional emission
below $\nu\lesssim 30\,$GHz is completely featureless.
}
\keywords{cosmic microwave background -- spectral distortions}

\maketitle

\section{Introduction}
With the advent of accurate observations of the Cosmic Microwave Background
(CMB) temperature and polarization anisotropies it becomes increasingly
important to understand the dynamics of cosmological recombination on the
percent level precision. Several authors \citep{Leung2004, Dubrovich2005,
Chluba2006, Kholu2006} have discussed physical processes leading to percent
level corrections to the results of the standard computation
\citep{Seager2000} for the ionization history. Recently, it has been shown
that also the treatment of the populations in the angular momentum sub-states
of hydrogen has percent level impact on the ionization history, especially at
redshifts $z\lesssim 800-1000$ \citep[][hereafter RMCS06 and CRMS06
respectively]{Jose2006, Chluba2006b} and it is expected that more physical
processes altering the dynamics of cosmological recombination may be realized.
\citet{Lewis2006} have made some first steps towards quantifying the possible
impact of percent level corrections to the ionization history on the estimation
of cosmological parameters.

In the future it may become possible to directly observe the spectral
distortions of the CMB arising during the epoch of recombination. This would
in principle open an alternative way to determine cosmological parameters like
the baryon and total matter density.
Several authors have discussed the distortions arising due to the hydrogen
higher level bound-bound transitions \citep[][RMCS06,
CRMS06]{Dubrovich1975,Liubarskii83,Rybicki93,DubroVlad95,Dubrovich2004,Kholu2005,Wong2006}.

In the standard computations within the cosmological context the {\it direct
recombination} to the ground state of hydrogen is neglected, since this
transition is so optically thick during the whole epoch of recombination that
escape of photons in the Lyman-continuum is considered impossible
\citep{Zeldovich68,Peebles68}.
On the other hand electrons can reach {\it any} of the other levels (with
$n\geq 2$) by {\it direct recombination} and the emitted continuum photons
escape freely, just because after their creation they will not encounter
another optically thick transition within the hydrogen atom.

Every successful recombination (i.e. when the electron is reaching the
ground-state and remains there) will therefore lead to the emission of {\it at
least} two photons, one from the direct recombination and (because direct
recombinations to the ground-state are neglected) {\it at least} one within
the cascade to the ground-state.
Here we show that for hydrogen a total of $\sim 4$ photons per neutral
hydrogen atom is produced within the bound-bound cascade and 2s two-photon
decay channel. Therefore, one expects a $\sim 25\%$ addition to the total
number of emitted photons during the epoch of recombination due to the {\it
direct recombination} to states with $n\geq 2$.

In this paper we discuss the hydrogen recombination spectrum in the frequency
range $100\,\text{MHz}\leq\nu\leq 3000\,\text{GHz}$, with special emphasis on
the contribution from the direct recombination lines (for more details on the
bound-bound emission within this context see RMCS06 and CRMS06). We use the
solution for the recombination history and evolution of the hydrogen
populations as obtained in CRMS06. In those computations a maximum of 100
shells following the populations of {\it all} the angular momentum sub-states
separately and including $l$- and $n$-changing collisions was treated.
We refer the reader to this paper for more details on the computations.

\section{Spectral distortion due to direct recombination}
In order to compute the spectral distortion arising due to direct
recombinations to a given level $(n,l)$ one has to consider the emission and
absorption of photons due to this process.
The the effective rate for the change of the population, $N_i$, of a hydrogen
level $i$ due to recombinations from the continuum is given by
\begin{equation}
\label{eq:DRic}
\left.\pAb{N_i}{t}\right|^{\rm rec}_{i} =\Ne N_{\rm p} R_{{\rm c}i}-N_i R_{i{\rm c}}
\Abst{,}
\end{equation}
where $\Ne$ and $N_{\rm p}$ are the free electron and proton number densities,
$R_{{\rm c}i}$ is the corresponding recombination rate and $R_{i{\rm c}}$ the
photoionization rate.
Using the definition of $R_{{\rm c}i}$ and $R_{i{\rm c}}$ in terms of the
photoionization cross section, $\sigma_{i{\rm c}}(\nu)$, one can write the
effective change of the photon number density, $N_\nu=I_\nu/h\nu$, due to
direct recombination to level $i$ at frequency $\nu$ as
\begin{equation}
\label{eq:DNic}
\frac{1}{c}\left.\pAb{N_\nu}{t}\right|^{\rm rec}_{i} 
=\Ne N_{\rm p}\,f_i(\Te)\,
\sigma_{i{\rm c}} 
\left[\frac{2 \nu^2}{c^2}+N_\nu \right] e^{-\frac{h\nu}{\kB\Te}}
-N_i\,\sigma_{i{\rm c}} N_\nu 
\Abst{.}
\end{equation}
Here stimulated recombination is included. From the Saha-relation one has
$f_i(\Te)=\left(\frac{N_{i}}{\Ne\,N_{\rm p}}\right)^{\rm LTE}
\!\!=\frac{g_i}{2}\left(\frac{h^2}{2\pi\,\me\kB \Te}\right)^{3/2}
e^{E_i/\kB \Te}$,
where $E_i$ is the ionization energy, $\Te$ is the electron temperature, which
always is very close to the radiation temperature $\Tg=T_0(1+z)$ with
$T_0=2.725\,$K, and $g_i=2 (2l+1)$ is the statistical weight of the level.
It is clear from equations \eqref{eq:DRic} and \eqref{eq:DNic} that the total
change of the number of photons has to be identical to the number of
recombined electrons. 
%

\begin{figure*}
\centering 
\includegraphics[width=1.94\columnwidth]{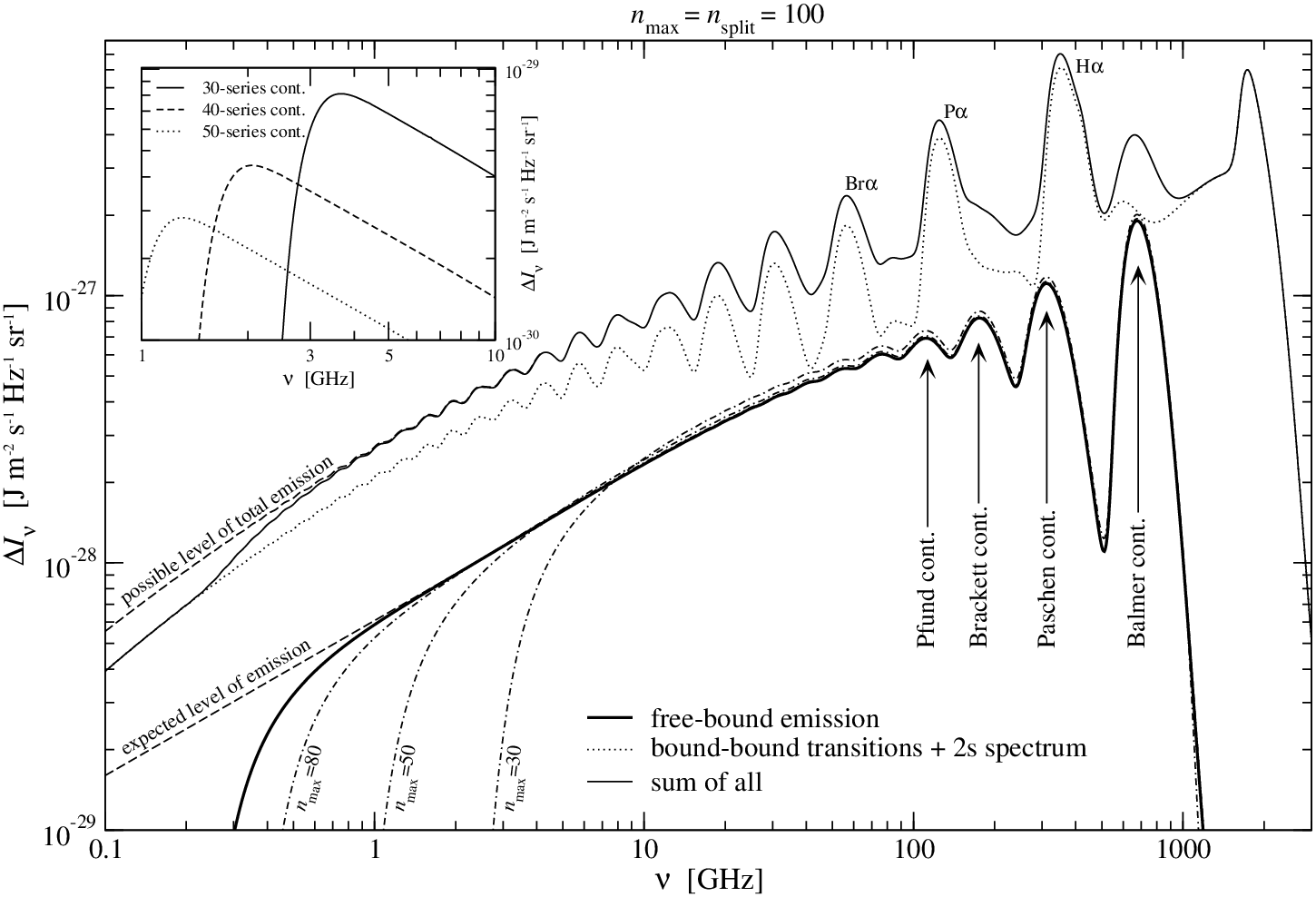}
\\[-0.0mm]
\includegraphics[width=1.94\columnwidth]{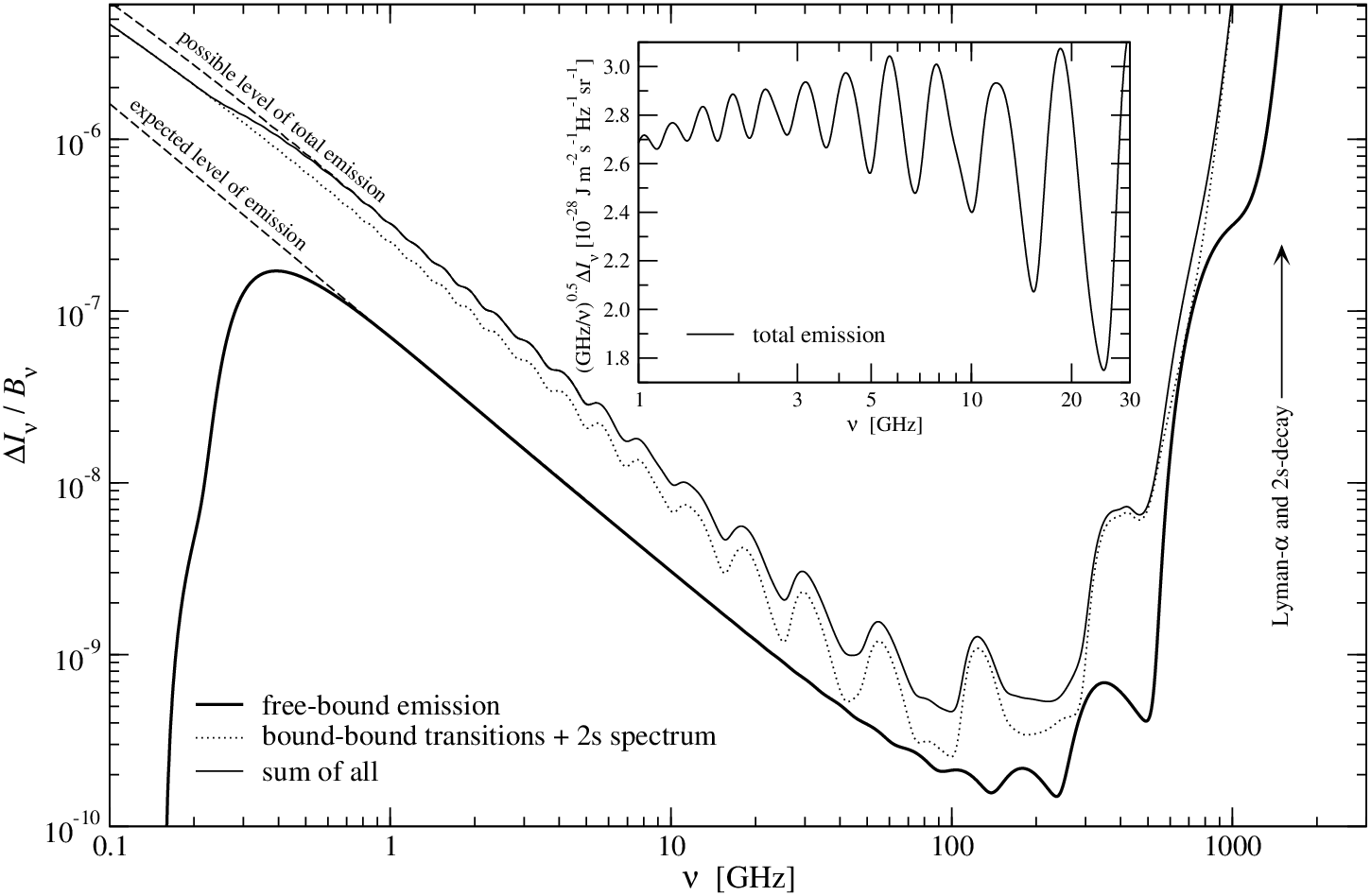}
\caption{The full hydrogen recombination spectrum including the free-bound
emission. 
The results of the computation for 100 shells as presented in CRMS06 were
used.
The contribution due to the 2s two-photon decay is also accounted for.
The dashed lines indicate the expected level of emission when including more
shells. In the upper panel we also show the free-bound continuum spectrum for
different values of $\nmax$ (dashed-dotted). The inlay gives the free-bound
emission for $n=30,\,40$ and $50$.
The lower panel shows the distortion relative to the CMB blackbody spectrum
and the inlay illustrates the modulation of the total emission spectrum
for $1\,\text{GHz}\leq \nu \leq 30\,\text{GHz}$ in convenient coordinates.}
\label{fig:DI_results}
\end{figure*}
Now, given the solution for the recombination history and evolution of the
populations $N_i$, one can obtain the solution for the change of the radiation
field in the optically thin limit assuming {\it a pure blackbody ambient
photon field}.
We find for the spectral distortion of the CMB at observing frequency $\nu$
due to direct recombination to level $i$ at redshift $z=0$
\begin{equation}
\label{eq:I_rec}
\Delta I^{\rm rec}_{i}(\nu)=
B_\nu
\!\!
\int_{z_{\rm t}}^\infty 
\!\!\!
\frac{c N_i \sigma_{i{\rm c}}(\nu_z)}{H(z)(1+z)}
\left[
\frac{\Ne\,N_{\rm p}}{N_{i}}\,f_i(\Te)\,e^{\frac{h\nu_z}{\kB\Tg}-\frac{h\nu_z}{\kB\Te}} -1
\right]
{\rm d} z
\Abst{,}
\end{equation}
with $\nu_z=\nu(1+z)$ and where $H(z)$ is the Hubble-expansion
factor, $1+z_{\rm t}=\nu_{i{\rm c}}/\nu$ corresponds to the redshift at
which the emission and photoionization threshold frequency, $\nu_{i{\rm c}}$,
are equal
and $B_\nu$ is the CMB blackbody spectrum today.

In the limit of $n\gg 1$ the ionization threshold corresponds to
$h\nu_n=\chi/n^2$ and the energy of the $n_\alpha$ transition is
$h\nu_{n_\alpha}=2\chi/n^3$, i.e. $n/2$ times smaller.
Therefore in our computations with $\nmax=100$ at $z=0$ the lowest
frequency we reach is expected to be of the order $\nu_{\rm low, fb}\sim
200-300\,$MHz instead of $\nu_{\rm low, bb}\sim 4-6\,$MHz for the bound-bound
transitions .
The width of the recombinational line is of the order $h\Delta\nu_n\sim \kB
T^{\rm peak}_{{\rm e},n}$, where $T^{\rm peak}_{{\rm e},n}\sim \Tg(z^{\rm
peak}_n)$ is the temperature of the electrons at the redshift, where the main
contribution to the {\it direct recombination emission} for shell $n$ appeared
(typically $z^{\rm peak}_n\sim 1300$).
Hence, one expects that for $h\nu_n\ll \kB T^{\rm peak}_{{\rm e},n}$ the
contributions to the free-bound continuum will become very broad. For the
Balmer and Paschen continuua one has $h\nu_2\sim 10\,\kB T^{\rm peak}_{{\rm
e},2}$ and $h\nu_3\sim 5\,\kB T^{\rm peak}_{{\rm e},3}$, so that the
contribution due to these transitions will be narrow, while for $n\gg 1$ they
will be very broad (see Fig.~\ref{fig:DI_results}).
%

\section{Results and discussion}

In Fig. \ref{fig:DI_results} we present the full hydrogen recombination
spectrum including the emission due to bound-bound transitions, the 2s
two-photon decay and direct recombinations to shells with $n\geq 2$. 
At high frequencies the free-bound emission shows narrow features
corresponding to the Balmer, Paschen, Brackett and Pfund continuua.
At low frequencies the free-bound continuua overlap very strongly and the
total continuum emission becomes completely featureless. 
The inlay plot in the upper panel shows the free-bound continuum emission for
three well separated high levels. These contributions are very broad
and each of them lies by a factor of $\sim 10$ below the total free-bound
emission spectrum. The total level of the emission is only reached after
summing over many shells.
As the inlay plot in the lower panel illustrates the sum of all contributions
(here mainly bound-bound and free-bound) still has a significant modulation
ranging from $\sim 4-35\%$ in the frequency band $1\,\text{GHz}\leq \nu \leq
30\,\text{GHz}$.
The slope of the free-bound distortion for $1\,\text{GHz}\lesssim \nu\lesssim
10\,$GHz is $\sim 0.6$ and therefore slightly steeper than for the
contribution from the bound-bound transitions alone ($\sim 0.46$). At
$\nu\lesssim 1\,$GHz one still expects an increase of emission when including
more than 100 shells.

In Fig. \ref{fig:DI_results} we also show the free-bound contribution for
different values of $\nmax< 100$. From this one can conclude that within our
assumptions the free-bound emission spectrum is converged to a level of better
than 1\% at $\nu\gtrsim 5-10\,$GHz and better than 10\% for
$1\,\text{GHz}\lesssim \nu\lesssim 2\,$GHz.
We have also checked how the free-bound emission is depending on the treatment
of the angular momentum sub-states and found that at high frequencies the
solution, when assuming full statistical equilibrium within the shells for
$n>2$, is not very different. At low frequencies, like in the case for
bound-bound transitions (see CRMS06), the difference is much larger, again
showing that it is important to follow the populations of all the angular
momentum sub-states separately.

\begin{figure}
\centering 
\includegraphics[width=0.94\columnwidth]{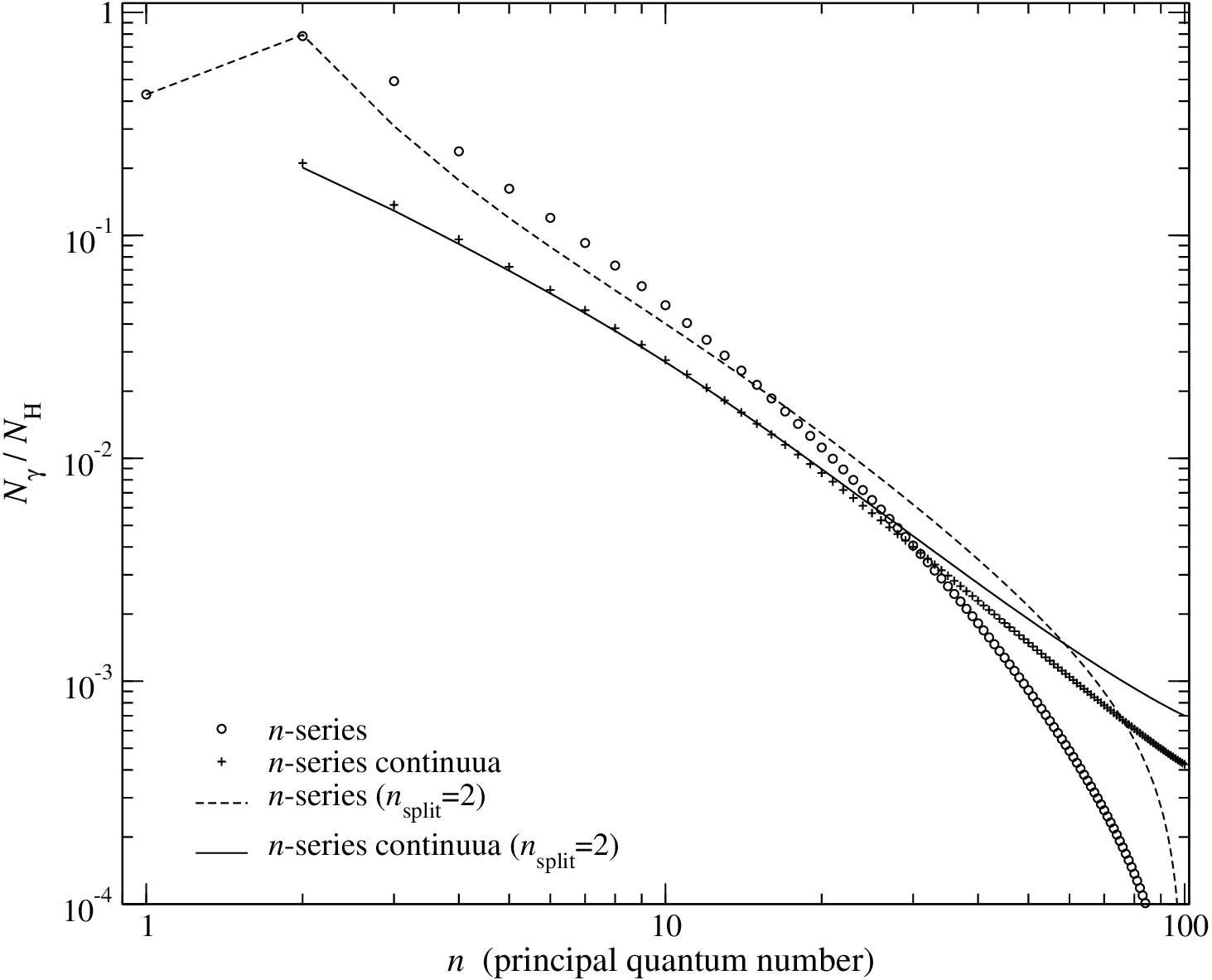}
\caption{Total number of photons per hydrogen atom released during
recombination for a given $n$-series and its continuum for $\nmax=100$. Also
the results assuming full statistical equilibrium within the shells for $n>2$
($\nsplit=2$) is presented.}
\label{fig:Photon_number}
\end{figure}

Figure~\ref{fig:Photon_number} shows the total number of photons per hydrogen
atom emitted during recombination for a given $n$-series (sum of all
bound-bound transitions to one shell with fixed $n$) and its continuum (the
corresponding bound-free contribution). In the considered case, one can see
that for $n\gtrsim n_{\rm cr} \sim 30$ the total emission is dominated by the
contribution from direct recombinations. For $n\lesssim n_{\rm cr}$ the
contribution from bound-bound-transitions dominates. This behavior shows that
for the lower shells cascading electrons are very important, whereas for
larger $n$ the electrons reach a given shell mainly via direct
recombinations. 
Obviously including more than 100 shells will lead to an increase of $n_{\rm
cr}$, but one does not expect a significant addition to the total number of
emitted photons.
There is a strong difference in both curves when the computation is done
assuming full statistical equilibrium for shell with $n>2$. In this case the
curves intersect at $n_{\rm cr}\sim 59$ instead of $n_{\rm cr} \sim 30$. This
shows that following all the angular momentum sub-states the lack of strong
redistribution to states with $l\gg 1$ disfavours the cascade with transitions
$\Delta n\ll n$.

Knowing the asymptotic behavior of the distortion at low frequencies, $\Delta
I_\nu=A_0\,[\nu/1\rm GHz]^\beta\, 10^{-28}\rm
J\,m^{-2}\,s^{-1}\,Hz^{-1}\,sr^{-1}$, from Fig.\ref{fig:DI_results}, one can
find the number of photons emitted per hydrogen atom (present density $N_{\rm
H}=\pot{1.9}{-7}\,\rm cm^{-3}$) at $\nu\leq\nu_0$, i.e. $N_\gamma/N_{\rm
H}=\frac{4\pi}{c N_{\rm H}}\int_0^{\nu_0} \frac{\Delta I_\nu}{h \nu}
\id\nu=\pot{3.3}{-2}\,\frac{A_0}{\beta}\,[\nu_0/1\rm GHz]^{\beta}$.
Within the range $1\,\text{GHz}\lesssim \nu\lesssim 10\,$GHz we obtained
$A_0^{\rm bb}\sim 2.2$ and $\beta_{\rm bb}\sim 0.46$ for the bound-bound and
$A_0^{\rm fb}\sim 0.59$ and $\beta_{\rm fb}\sim 0.6$ for the free-bound
distortion.
Considering the bound-bound distortion assuming that the $\alpha$-transitions
give the main contribution and that most of the photons are released at
$z_{\rm em}\sim 1300$, with $h\nu_{n_\alpha}^{\rm obs}=2\chi/z_{\rm em}\,n^3$,
one finds
$N^{\rm bb}_\gamma(n>n_0)/N_{\rm H}=\pot{3.3}{-2}[\pot{6.6}{6}/z_{\rm
em}]^\beta\,\frac{A_0}{\beta}\,n_0^{-3\beta}\sim 8.0\,n_0^{-1.38}$
and $|\id N^{\rm bb}_\gamma/\id n_0|\sim 11\,n_0^{-2.38}\,N_{\rm H}$ for the
contribution of the whole series for given shell $n_0$.
Similarly, for the free-bound emission with $h\nu_{n}^{\rm obs}=\chi/z_{\rm
em}\,n^2$ one has $N^{\rm fb}_\gamma(n>n_0)/N_{\rm H}\sim 3.6\,n_0^{-1.2}$ and
$|\id N^{\rm fb}_\gamma/\id n_0|\sim 4.3\,n_0^{-2.2}\,N_{\rm H}$, although for
$n_0\gg 1$ due to the large width of the free-bound contributions one expects
this estimate to be rather crude.
We found that $|\id N^{\rm fb}_\gamma/\id n_0|\sim 2.5\,n_0^{-1.9}\,N_{\rm H}$
represents the result shown in Fig.~\ref{fig:Photon_number} for large $n_0$
very well.
One should mention that for $n\gg 1$ and $|\Delta n|\ll n$ induced
recombinations and stimulated transitions play an important role.

\begin{table}
\caption{Total number of photons per hydrogen atom emitted during
recombination within a given series and its continuum. Adding all the
contributions one obtains $4.98$ photons per hydrogen atom.}
\label{tab:photon_number}
\centering
\begin{tabular}{@{}lcccc}
\hline
\hline
Series & $n$ & bound-bound & continuum & sum \\
\hline
Lyman & 1  & $\pot{4.28}{-1}$ & 0  & $\pot{4.28}{-1}$ \\
Balmer & 2  & $\pot{7.83}{-1}$  &  $\pot{2.11}{-1}$ & $\pot{9.94}{-1}$ \\
Paschen & 3  & $\pot{4.91}{-1}$  &  $\pot{1.37}{-1}$ & $\pot{6.28}{-1}$ \\
Brackett & 4  & $\pot{2.38}{-1}$  &  $\pot{9.59}{-2}$ & $\pot{3.34}{-1}$ \\
Pfund & 5  & $\pot{1.62}{-1}$  &  $\pot{7.22}{-2}$ & $\pot{2.34}{-1}$ \\
All & 1-100  & 2.84  &  1.00 & 3.84 \\
\hline
\hline
2s-decay & 1  & 0 & 1.14  & 1.14 \\
\end{tabular}
\end{table}
In Table~\ref{tab:photon_number} we give the values of the total number of
photons per hydrogen atom emitted in the Balmer, Paschen, Brackett and
Pfund-series and their continuua. Also we listed the total number of emitted
photons in all $n$-series and their continuua and the contribution from the 2s
two-photon decay.
As expected the total number of emitted photons due to free-bound transitions
(within the accuracy of our calculation) is identical to the total number of
hydrogen atoms. The number of photons emitted within the Balmer-series and
Balmer-continuum is also very close to one photon per hydrogen atom.
One can see that roughly $2.8$ photons per hydrogen atom are released within
the bound-bound transitions.
For the free-bound distortion 80\% of the photon are emitted for $n\lesssim
14-15$, 90\% for $n\lesssim 26$ and 95\% for $n\lesssim 41$. On the other hand
within the bound-bound transitions one finds 80\% for $n\lesssim
6-7$, 90\% for $n\lesssim 11-12$ and 95\% for $n\lesssim 18$.
Focusing on the Lyman-series and the 2s two-photon decay contribution one can
see that $\sim 43\,\%$ of all electron go through the Lyman-$\alpha$ and $\sim
57\,\%$ of electron reach the ground state via the 2s two-photon decay
channel.
{\it During the epoch of hydrogen recombination a total of $\sim 5$ photons per
hydrogen atom are produced.}
Therefore, recombination of hydrogen slightly increases the specific entropy
of the Universe (photons per baryon).

Observations of the recombination features at frequencies $\nu\gtrsim
1412\,$MHz might become feasible since the strength of the signal under
discussion is close to $\pot{2}{-7}B_\nu$ and still has variability with well
defined frequency dependence on the level of several percent
(Fig.~\ref{fig:DI_results}). Also in this band ($\lambda<21\,$cm) one does not
expect other sources with similar frequency dependence.

\acknowledgement{ 
We wish to thank J.A. Rubi\~no-Mart\'{\i}n for useful
discussion on this problem at the initial stage.  
}

\bibliographystyle{aa}
\bibliography{Lit}

\begin{thebibliography}{17}
\expandafter\ifx\csname natexlab\endcsname\relax\def\natexlab#1{#1}\fi

\bibitem[{{Chluba} {et~al.}(2006){Chluba}, {Rubi\~no-Martin}, \&
  {Sunyaev}}]{Chluba2006b}
{Chluba}, J., {Rubi\~no-Martin}, J.~A., \& {Sunyaev}, R.~A. 2006, in
  preparation, CRMS06

\bibitem[{{Chluba} \& {Sunyaev}(2006)}]{Chluba2006}
{Chluba}, J. \& {Sunyaev}, R.~A. 2006, \aap, 446, 39

\bibitem[{{Dubrovich}(1975)}]{Dubrovich1975}
{Dubrovich}, V.~K. 1975, Soviet Astronomy Letters, 1, 196

\bibitem[{{Dubrovich} \& {Grachev}(2005)}]{Dubrovich2005}
{Dubrovich}, V.~K. \& {Grachev}, S.~I. 2005, Astronomy Letters, 31, 359

\bibitem[{{Dubrovich} \& {Shakhvorostova}(2004)}]{Dubrovich2004}
{Dubrovich}, V.~K. \& {Shakhvorostova}, N.~N. 2004, Astronomy Letters, 30, 509

\bibitem[{{Dubrovich} \& {Stolyarov}(1995)}]{DubroVlad95}
{Dubrovich}, V.~K. \& {Stolyarov}, V.~A. 1995, \aap, 302, 635

\bibitem[{Kholupenko \& Ivanchik(2006)}]{Kholu2006}
Kholupenko, E.~E. \& Ivanchik, A.~V. 2006, submitted to Astronomy Letters

\bibitem[{Kholupenko {et~al.}(2005)Kholupenko, Ivanchik, \&
  Varshalovich}]{Kholu2005}
Kholupenko, E.~E., Ivanchik, A.~V., \& Varshalovich, D.~A. 2005,
  arXiv:astro-ph/0509807

\bibitem[{{Leung} {et~al.}(2004){Leung}, {Chan}, \& {Chu}}]{Leung2004}
{Leung}, P.~K., {Chan}, C.~W., \& {Chu}, M.-C. 2004, \mnras, 349, 632

\bibitem[{{Lewis} {et~al.}(2006){Lewis}, {Weller}, \& {Battye}}]{Lewis2006}
{Lewis}, A., {Weller}, J., \& {Battye}, R. 2006, astro-ph/0606552

\bibitem[{{Liubarskii} \& {Sunyaev}(1983)}]{Liubarskii83}
{Liubarskii}, I.~E. \& {Sunyaev}, R.~A. 1983, \aap, 123, 171

\bibitem[{{Peebles}(1968)}]{Peebles68}
{Peebles}, P.~J.~E. 1968, \apj, 153, 1

\bibitem[{{Rubi\~no-Martin} {et~al.}(2006){Rubi\~no-Martin}, {Chluba}, \&
  {Sunyaev}}]{Jose2006}
{Rubi\~no-Martin}, J.~A., {Chluba}, J., \& {Sunyaev}, R.~A. 2006, submitted to
  \mnras, astro-ph/0607373, RMCS06

\bibitem[{{Rybicki} \& {dell'Antonio}(1993)}]{Rybicki93}
{Rybicki}, G.~B. \& {dell'Antonio}, I.~P. 1993, in ASP Conf. Ser. 51:
  Observational Cosmology, ed. G.~L. {Chincarini}, A.~{Iovino}, T.~{Maccacaro},
  \& D.~{Maccagni}, 548--+

\bibitem[{{Seager} {et~al.}(2000){Seager}, {Sasselov}, \& {Scott}}]{Seager2000}
{Seager}, S., {Sasselov}, D.~D., \& {Scott}, D. 2000, \apjs, 128, 407

\bibitem[{{Wong} {et~al.}(2006){Wong}, {Seager}, \& {Scott}}]{Wong2006}
{Wong}, W.~Y., {Seager}, S., \& {Scott}, D. 2006, \mnras, 367, 1666

\bibitem[{{Zeldovich} {et~al.}(1968){Zeldovich}, {Kurt}, \&
  {Syunyaev}}]{Zeldovich68}
{Zeldovich}, Y.~B., {Kurt}, V.~G., \& {Syunyaev}, R.~A. 1968, Zhurnal
  Eksperimental noi i Teoreticheskoi Fiziki, 55, 278

\end{thebibliography}

\end{document}